\documentclass[final, times, 5p, twocolumn]{elsarticle}
\usepackage{hyperref}
\usepackage{float}
\usepackage{caption}
\usepackage{lineno}
\usepackage{enumitem}
\usepackage{subfigure}
\usepackage{listings}
\usepackage[algoruled,vlined]{algorithm2e}
\usepackage{color}
\SetSideCommentRight

\journal{Computer Physics Communicaions}

\DeclareFixedFont{\ttb}{T1}{txtt}{bx}{n}{12} % for bold
\DeclareFixedFont{\ttm}{T1}{txtt}{m}{n}{12}  % for normal

\usepackage{xcolor}
\definecolor{deepblue}{rgb}{0,0,0.5}
\definecolor{Jesse}{rgb}{0,0.608,0.467}
\definecolor{deepred}{rgb}{0.6,0,0}
\definecolor{deepgreen}{rgb}{0,0.5,0}
\definecolor{lightskyblue}{rgb}{0.53, 0.81, 0.98}
\definecolor{lightcyan}{rgb}{0.88, 1.0, 1.0}

\def\StartLineAt#1{\lstset{firstnumber=#1}}

\newcommand\pythonstyle{\lstset{
language=Python,
basicstyle=\ttm\footnotesize,
otherkeywords={self},             % Add keywords here
keywordstyle=\ttb\footnotesize\color{deepblue},
emph={MyClass,__init__},          % Custom highlighting
xleftmargin=0.3cm,
keepspaces=true,
emphstyle=\ttb\footnotesize\color{deepred},    % Custom highlighting style
stringstyle=\footnotesize\color{deepgreen},
showstringspaces=false            % 
}}

\lstnewenvironment{python}[1][]
{
\pythonstyle
\lstset{#1}
}
{}

\newcommand\pythoninline[1]{{\pythonstyle\lstinline!#1!}}

\usepackage{xcolor}
\usepackage{amsmath}

\SetCommentSty{xCommentSty}
\DeclareMathOperator*{\argmin}{argmin}
\newlength\algowd

\SetAlFnt{\small}
  
\begin{document}

\begin{frontmatter}

\title{\textit{NanoNET}: an extendable Python framework for semi-empirical tight-binding models}

\author[exciton,chemphys]{M. V. Klymenko\corref{mycorrespondingauthor}}
\ead{mike.klymenko@rmit.edu.au}
\author[fleet,chemphys]{J. A. Vaitkus}
\author[fleet,chemphys]{J. S. Smith}
\author[exciton,fleet,chemphys]{J. H. Cole}
\cortext[mycorrespondingauthor]{Corresponding author}

\address[exciton]{ARC Centre of Excellence in Exciton Science, RMIT University, Melbourne, Victoria 3001, Australia}
\address[fleet]{ARC Centre of Excellence in Future Low-Energy Electronics Technologies, RMIT University, Melbourne, Victoria 3001, Australia}
\address[chemphys]{Chemical and Quantum Physics, School of Science, RMIT University, Melbourne, Victoria 3001, Australia }

\begin{abstract}
We present a novel open-source \emph{Python} framework called \textit{NanoNET} (Nanoscale Non-equilibrium Electron Transport) for modelling electronic structure and transport. Our method is based on the tight-binding method and non-equilibrium Green's function theory. The core functionality of the framework is providing facilities for efficient construction of tight-binding Hamiltonian matrices from a list of atomic coordinates and a lookup table of the two-center integrals in dense, sparse, or block-tridiagonal forms. The framework implements a method based on $kd$-tree nearest-neighbour search and is applicable to isolated atomic clusters and periodic structures. A set of subroutines for detecting the block-tridiagonal structure of a Hamiltonian matrix and splitting it into series of diagonal and off-diagonal blocks is based on a new greedy algorithm with recursion. Additionally the developed software is equipped with a set of programs for computing complex band structure, self-energies of elastic scattering processes, and Green's functions. Examples of usage and capabilities of the computational framework are illustrated by computing the band structure and transport properties of a silicon nanowire as well as the band structure of bulk bismuth.
\end{abstract}

\begin{keyword}
tight-binding method \sep Hamiltonian matrix \sep $kd$-tree \sep band matrix \sep block-tridiagonal matrix \sep non-equilibrium Green's functions
% \MSC[2010] 00-01\sep  99-00
\end{keyword}

\end{frontmatter}

{\bf PROGRAM SUMMARY}
  %Delete as appropriate.

\begin{small}
\noindent
{\em Program Title: NanoNET}                                          \\
{\em Developer's repository link: https://github.com/freude/NanoNet}                                          \\
{\em Licensing provisions: MIT}                                       \\
{\em Programming language: Python}                                    \\

{\em Nature of problem: }\\
The framework NanoNET solves a problem which is, having a set of atomic coordinates and tight-binding parameters, to construct Hamiltonian matrices in one of several desired forms. In particular, some applications require those matrices to have a reduced bandwidth and/or to possess a block-tridiagonal structure.\\

{\em Solution method:}\\
  The problem is solved using a combination of $kd$-tree-based fast nearest-neighbour search and atomic coordinate sorting. Furthermore, a new greedy recursive algorithm is proposed for detecting block-tridiagonal structure of a matrix in a non-optimal way. Additionally, we propose an algorithm of a polynomial time for optimizing block sizes.\\
  
{\em Additional features: }\\
    Although the resulting matrices can be processed by many existing software packages, the framework also has built-in standard tools for diagonalizing Hamiltonian matrices and computing Green's functions that make it an independent tool for solving electronic structure and transport problems.
   \\
\end{small}

% \linenumbers

\section{Introduction}
The tight-binding (TB) method coupled with the non-equilibrium Green's function (NEGF) formalism is a widely used method for simulations of electronic devices at the atomic level \cite{Neophytou2008} including large-scale FinFETs \cite{Lansbergen2008}, nanowire FETs \cite{PhysRevB.73.165319, Zheng, Afzalian_2018}, single-atom transistors \cite{Ryu, Smith2015}, etc. The TB method is a method to tackle large-scale electronic structure  problems \cite{Rudenko,PhysRevB.58.7260, Elstner, Calogero_2018} by both limiting the size of a basis set and taking into account only interactions between a finite number of neighbouring atoms. As a result, even for large numbers of atoms, the Hamiltonian matrix of a system in the TB representation is sparse. Computing the inverse and eigenvalues of this matrix scales almost linearly with the system size using approximate numerical technique like Krylov subspace methods \cite{Bowler_1997, PhysRevB.51.10157, PhysRevB.53.12733,Elstner, Calogero_2018}. NEGF theory, often taking TB Hamiltonians as an input, provides a set of tools to solve electron transport problems for systems with both elastic and inelastic scatterings \cite{Bonitz, Stefanucci, PhysRevB.97.085149}, spatially inhomogeneous parameters and open boundary conditions \cite{Datta, negf}.

The Hamiltonian operator in the tight-binding representation may be written in the form:
\begin{equation}
    \hat{H} = \sum_{m,n} H_{n,m} c_n^{\dagger} c_m
\end{equation}
where $H_{n,m}$ is an element of the Hamiltonian matrix $H$, $c_n^{\dagger}$ and $c_m$ are the creation and annihilation operators respectively, and indices $n$ and $m$ identify basis functions. In the tight-binding representation each of the indices correspond to a set consisting of a position vector of an atomic site, and a set of quantum numbers associated with a localized basis set for each site. The algorithms presented in this work are aimed at efficiently constructing the matrix $H_{n,m}$ given a list of atomic coordinates and a lookup table of two-center integrals as input parameters. Note that a matrix representation of the Hamiltonian using a linear combination of atomic orbitals (LCAO) is not unique and can be obtained in various ways, resulting in matrices with varying sparsity patterns depending on the order at which atoms are addressed when the Hamiltonian matrix is built (see Fig. \ref{fig:intro} for an illustrative example). Although permutations of atomic coordinates may lead to physically equivalent matrix representations of the Hamiltonian, some of them work better with certain numerical methods resulting in faster computations. 

There are several ways to extract physically relevant information from the Hamiltonian. For isolated systems in thermodynamic equilibrium, one may compute properties by diagonalizing the Hamiltonian matrix. For the systems with open boundary conditions, one computes the so-called Green's function matrix, $G$  \cite{negf}:
\begin{equation}
    G = \left[ (E+i0^+) I - H - \Sigma \right]^{-1}
\end{equation}
where $E$ is the energy, $0^+$ is a small positive infinitesimal, $H$ is the Hamiltonian and $\Sigma$ is the self-energy describing an exchange of electrons through open boundaries. In general, a numerically exact computation of the matrix inverse scales as $\mathcal{O}(N^3)$ with the number of diagonal elements $N$. Better scalablity may be achieved exploiting the sparsity of the matrix and taking into account that only some but not all matrix elements of the inverse are of interest for physical applications \cite{negf}. There are two main strategies for this. One is to construct a hierarchical method such as the method developed by Ozaki \cite{Ozaki} using a set of formulae derived from a recursive application of a block $LDL^T$ factorization using the Schur complement to a structured matrix obtained by a nested dissection for the sparse matrix. The method scales as $O[N\left(\log_2 N\right)^2]$, $O\left(N^2 \right)$, and $O\left(N^{7/3} \right)$ for one-, two-, and three-dimensional systems respectively. The second strategy is to minimize the bandwidth of the matrix and to utilize its block-tridiagonal structure to improve the computational efficiency by using the recursive Green's function (RGF) algorithm to compute the inverse as a sequence of layers, scaling as $\mathcal{O}(\sum_j^k N_j^3)$, where $k$ is the number of diagonal blocks and $N_j$ is the number of diagonal elements in the $j^\text{th}$ block. If all blocks are equal in size this corresponds to a $k^2$ improvement in both memory and run-time. These examples motivate development of a software framework that provides facilities to construct TB Hamiltonians in different formats defined by a particular application.

In this work, we propose a method to construct TB Hamiltonian matrices in several possible forms (with different sparsity patterns) including dense, sparse, and block-tridiagonal matrices. Section 2 describes algorithms providing facilities for efficient building of such matrices. Section 3 presents the software architecture of \textit{NanoNET} consisting of two Python packages \texttt{tb} and \texttt{negf}: the first one implements new algorithms for composing TB matrices, and the latter contains tools for computing NEGF and other physical quantities related to the electron transport. Relationships between TB matrices, NEGF, and other quantities of interest are briefly listed in Section 4. Section 5 contains several illustrative examples of \textit{NanoNET} usage.

\emph{NanoNET} is an open-source software providing a framework for electron transport computations based on the TB method. Although its functionality somewhat overlaps with other software such as \emph{NEMO5} \cite{NEMO}, \emph{Kwant} \cite{kwant}, \emph{pybinding} \cite{pybinding}, \emph{Transiesta} \cite{TransSIESTA}, \emph{Smeagol} \cite{Smeagol}, \emph{Gollum} \cite{Gollum}, \emph{DFTB+} and combined \emph{ASE} \cite{ase} and \emph{GPAW} \cite{gpaw}, the framework has following distinctive features: it provides tools for processing and outputting TB matrices with desired properties, it focuses on the semi-empirical tight-binding method with a flexible setting of tight-binding parameters together with coordinate representations of an atomic structure, and it provides a \emph{Python} application programming interface to NEGF computations.

A framework architecture can be seen as a set of more or less independent building blocks and pipeline schemes that can be used in a user-written code. This flexibility allows the user to pick algorithms that suit best for his particular problem to achieve the best performance. Due to its compact syntax, \emph{Python} seems to be a particularly convenient programming language for implementing programming interfaces for the software frameworks (examples of popular Python frameworks are \emph{Tensorflow} from \emph{Google} \cite{tensorflow} for building neural networks, \emph{Flask} for web development etc.). For the electronic transport problems, examples of an existing software framework are \emph{Kwant} and the stack \emph{ASE+GPAW}. In these frameworks, the user has multiple choices of computational methods at each stage of the calculations as well as the flexibility to modify dataflow in many possible ways, which is useful for building software interfaces and post-processing. Comparing to \emph{Kwant}, our framework is more oriented on the processing of atomic coordinates, while, comparing to the \emph{ASE+GPAW} stack, we provide a tool to work with semi-empirical tight-binding parameters and detecting the tri-blockdiagonal structure of matrices.

The framework interacts with a user via a command line interface or \emph{Python} application programming interface inspired by the \emph{ASE} \cite{ase} and \emph{GPAW} \cite{gpaw} software packages. In future we will provide additional interfaces to achieve compatibility with \emph{ASE} and \emph{Kwant} in order to enhance mutual functionality of these frameworks.

\section{Algorithms}

\subsection{Construction of Hamiltonian matrices}

 Using a list of atomic coordinates, $L$, the Hamiltonian matrix $H$ is constructed in the framework of the TB method via Algorithm \ref{a1}. Each entry in $L$ contains information of an atomic site $site_j$; the numerical index of the site, $j$, the name of chemical element, $label$, and Cartesian coordinates, $coords$. The label is used to search a look-up table, $OrbitalsTable$, which associates a list of orbitals with each of the chemical elements.
 
 The algorithm starts with one of the sorting procedures defined by a user or embedded in the framework. Different matrix representations are determined by a particular order of atomic coordinates which are in turn computed by a user-defined sorting procedure or one of the embedded sorting algorithms (see Section 2.4 for more details). Typically sorting algorithms scale as $\mathcal{O}(N \log N)$ with the number of atoms. The location of matrix elements in the resulting matrix can be quickly determined using a pre-computed set of index offsets, even though each site can possess a different number of orbitals (see lines 3--5, 10 and 13 of Algorithm \ref{a1}). When the order of atoms is determined, the Hamiltonian matrix element can be computed by addressing each atom in sequence and finding its nearest neighbours with which the atom has non-zero coupling. In order to speed up the nearest neighbour search procedure, the atomic coordinates are arranged in a $kd$-tree --- a space-partitioning data structure for organizing points in a $k$-dimensional space \cite{kd-tree, Maneewongvatana}. The nearest neighbours search itself is a querying of the constructed tree. Building the tree scales as $\mathcal{O}(kN \log N)$, though one query can be done in $\mathcal{O}(\log N)$ time. The total time of the nearest-neighbour search for all atoms scales as $\mathcal{O}(N \log N)$. After arranging all atomic coordinates into a $kd$-tree, Algorithm \ref{a1} loops over all atoms (line 7 of Algorithm \ref{a1}) and their orbitals (line 9), finds nearest neighbours for each site (line 8), loops over orbitals of the neighbours (lines 11 and 12), and computes the matrix elements of $H$ (line 14).  Note that in such an implementation, the algorithm does not run over pairs of atoms that are not coupled. Therefore each time the algorithm invokes the function \texttt{get\_matrix\_element()}, it most likely returns a non-zero value if the coupling is allowed by orbital symmetry. As a result, the Hamiltonian matrix can be \textit{directly} constructed in one of several existing sparse matrix formats, since only non-zero elements are processed and their indices, $j_1$ and $j_2$, are explicitly computed.
 
 %%%%%%%%%%%%%%%%%%%%%%%%%%%%%%%%%%%%%%%%%%%%%%%%%%%%%%%%%%%%%%%%%%%%%%%%%%%%%%%%%

\begin{algorithm}[!h]
\SetKwComment{com}{}{}
\SetKwFunction{neighbours}{get\_neighbours}
\SetKwFunction{me}{get\_matrix\_element}
\SetKwFunction{index}{get\_index}
\SetKwFunction{init}{initialize\_Hamiltonian}
\SetKwFunction{kd}{make\_kd\_tree}
\SetKwFunction{sort}{sort}
\SetKwFunction{len}{size}
\SetSideCommentRight
\DontPrintSemicolon
\caption{\label{a1} Computing Hamiltonian matrix for a cluster of atoms}
\Begin{
let $L$ is a list of atomic coordinates \;
\tcp{Sort the list of atomic coordinates, $ \mathcal{O}(N \log N) $}
$L_s \leftarrow \sort(L)$\;% \tcp*{Sort the list of atomic coordinates, $ \mathcal{O}(N \log N) $}
% \tcp{Compute index offsets defining positions of site indices in the Hamiltonian matrix}
\tcp{Compute index offsets}
$Offsets[1] = 0$ \;
\For{$j = 2,...,N$}{
    $\textit{Offsets}[j] \leftarrow \textit{Offsets}[j-1] +$\\ \hspace{1.3cm} $\len(OrbitalsTable[L_s[j].label]) $\;%  \tcp*{Compute index offsets}
}
\tcp{Build $kd$-tree, $\mathcal{O}(3N \log N) $}
$kd\_tree \leftarrow \kd(L_s[1,...,N].coords)$\;% \tcp*{Build $kd$-tree, $\mathcal{O}(3N \log N) $}
\tcp{Compute TB Hamiltonian}
\tcp{Loop over sites}
\For{$site_1 \in L_s$}{
    \tcp{Find nearest neighbours for $site_1$, $\mathcal{O}(\log N)$ }
    $L_n \leftarrow kd\_tree.\neighbours(site_1)$\;%\tcp*{Find nearest neighbours for $site_1$, $\mathcal{O}(\log N)$ }
    \tcp{Loop over orbitals of $site_1$}
    \For{$orbital_1 \in OrbitalsTable[site_1.label]$}{
        $j_1 \leftarrow \textit{Offsets}[site_1.j] + orbital_1.index$\;
    \tcp{Loop over neighbours of $site_1$}
    \For{$site_2 \in L_n$}{
            \tcp{Loop over orbitals of $site_2$}
            \For{$orbital_2 \in OrbitalsTable[site_2.label]$}{
                $j_2 \leftarrow \text{Offsets}[site_2.j] + orbital_2.index$\;
                \tcp{Compute matrix element}
                $H[j_1, j_2] \leftarrow \me(site_1,$\\ \hspace{1.3cm} $orbital_1,\ site_2,\ orbital_2)$ \tcp*{$\mathcal{O}(1)$}
            }
        }
    }
}
}
\end{algorithm}

\begin{algorithm}[!h]
\SetKwComment{com}{}{}
\SetKwFunction{neighbours}{get\_neighbours}
\SetKwFunction{me}{get\_matrix\_element}
\SetKwFunction{tr}{translate}
\SetKwFunction{init}{initialize\_Hamiltonian}
\SetKwFunction{kd}{make\_kd\_tree}
\SetKwFunction{sort}{sort}
\SetKwFunction{append}{append}
\SetSideCommentRight
\DontPrintSemicolon
\caption{\label{a2}Computing coupling Hamiltonians for periodically translated cells}
\Begin{
let $T[1,...,D]$ be translation vectors \;
\tcp{Create a list of the translated atomic coordinates}
$L_s^t \leftarrow \text{empty list}$ \;
\For{$vector \in T$}{
    \For{$site \in L_s$}{
        $L_s^t.\append(\tr(site, vector))$
    }
}
\tcp{Build new $kd$-tree for the neighbours outside the unit cell}
$kd\_tree^t \leftarrow \kd(L_s^t)$ \;
\tcp{Compute the inter-cell coupling Hamiltonian, $H_C$}
\For{$site_1 \in L_s$}{
    $L_n \leftarrow kd\_tree^t.\neighbours(site_1)$ \;
    \For{$orbital_1 \in OrbitalsTable[site_1.label]$}{
        $j_1 \leftarrow \textit{Offsets}[site_1.j] + orbital_1.index$\;
    \For{$site_2 \in L_n$}{
            \For{$orbital_2 \in OrbitalsTable[site_2.label]$}{
                $s = site_2.origin$ \;
                $j_2 \leftarrow \textit{Offsets}[s] + orbital_2.index$\;
                $H_C[j_1, j_2] \leftarrow  \me(site_1,$\\ \hspace{1.3cm} $orbital_1,\ site_2,\ orbital_2)$
            }
        }
    }
}
}
\end{algorithm}

%%%%%%%%%%%%%%%%%%%%%%%%%%%%%%%%%%%%%%%%%%%%%%%%%%%%%%%%%%%%%%%%%%%%%%%%%%%%%

\begin{figure}[t]
    \centering
    \includegraphics{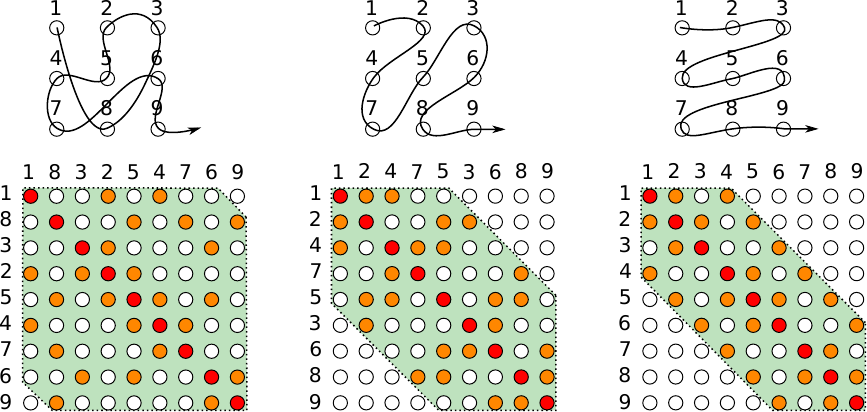}
    \caption{Dependence of the matrix bandwidth on the order at which rows of the Hamiltonian matrix are computed}
    \label{fig:intro}
\end{figure}

\subsection{Periodic boundary conditions and coupling between unit cells}

The periodic boundary conditions are determined by a set of translation vectors $T$ defining a primitive cell of the periodic structure. The Hamiltonian describing coupling between primitive cells $H_C$ is computed according to Algorithm \ref{a2}. The algorithm generates the list of atomic coordinates $L_s^t$ by acting translation operators on the coordinates of each atom inside the primitive cell (line 6) in addition to the list of atomic coordinates $L$ in  Algorithm \ref{a1}. For large structures, the computational burden can be reduced by translating solely the atoms adjacent to the surfaces of the primitive cell.  Comparing to the list $L$, the entries of the list $L_s^t$ have one additional field, $origin$, which is the index of the original atom being a pre-image of an atom after the translation. The rest of the steps of Algorithm \ref{a2} are similar to those of Algorithm \ref{a1} with the main distinction that the $kd$-tree is built only for the atoms outside the primitive cell and the matrix elements are computed for the pairs of atoms interacting across the borders of the primitive cell. Thus, when periodic boundary conditions are applied, two $kd$-trees are built: one is for atoms within a primitive cell and another one is for the atoms from neighboring/adjacent primitive cells. The matrix $H_C$ has same shape as $H$.

\subsection{Computing Hamiltonian matrix elements}

Usually in TB methods, the electronic structure problems are formulated in terms of two-center integrals neglecting three-center contributions. Computation of these integrals is performed by a function \texttt{get\_matrix\_element()} in Algorithms \ref{a1} and \ref{a2}. In the semi-empirical version of the TB method, two-center integrals depend on a relatively small number of empirical parameters which, according to Ref. \cite{Slater}, can be reduced to the two-center integrals of a diatomic molecule, designated by a tuple of three quantum numbers, e.g. $(ss\sigma)$, $(pp\pi)$ etc. Mapping those parameters to the two-center integrals is achieved via coordinate transformations from the diatomic molecule frame to the crystal frame. Explicit formulas for those transformations have been derived by Slater and Koster for many possible combinations of angular momentum quantum numbers \cite{Slater}. Since the number of the coordinate transformations is rather large, we have computed angular dependence of the diatomic two-center integrals using an approach proposed by Podolskiy and Vogt \cite{Podolskiy} instead of using a table of explicit formulas. Podolskiy and Vogt have derived compact close analytic expressions for the angular dependence of tight-binding Hamiltonian matrix
elements in the two-center approximation that are well suited for numerical calculations and are valid for all angular momentum quantum numbers.

In our codebase, the diatomic two-center integrals are introduced using meta-programming features of \emph{Python}. Namely, the parameters are arranged as a module-level variable created at runtime. Each variable is a Python dictionary with a certain naming convention (see Section 4) containing a set of parameters for coupling between a pair of elements. 

The matrix elements can be computed beyond the first nearest neighbour coupling if the corresponding TB parameters are provided. In \textit{NanoNET} one can specify two types of radial dependence functions: one, depending on the distance between atoms, outputs an integer number which allows for location of a certain set of tight-binding parameters (the number is used as a part of name convention and designates first, second, third etc. nearest neighbours), and the other outputs a float number which is then multiplied on the matrix element. If the radial dependence functions are not specified, \textit{NanoNET} will search for and use a single set of TB parameters. Also, if not all orbitals in the basis set have same spin quantum number and the spin-orbit coupling energy is set, the Hamiltonian matrix includes the spin-orbit coupling terms computed for p-orbitals.

\subsection{Sorting the list of atomic coordinates and computing block-tridiagonal representation of a matrix}

\begin{figure}[t]
    \centering
    \subfigure[$b=5953$]{\includegraphics{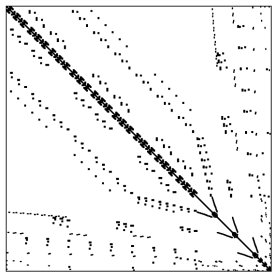}}
    \subfigure[$b=1091$]{\includegraphics{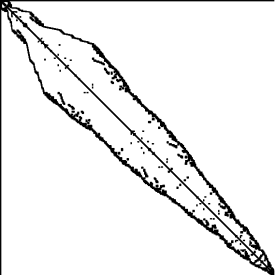}}
    \subfigure[$b=547$]{\includegraphics{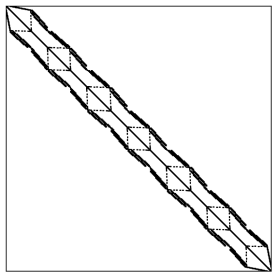}}
    \caption{Sparsity patterns and matrix bandwidth, $b$, for a) arbitrary ordering of atoms, b) after bandwidth reduction using reverse Cuthill-McKee algorithm, and c) after sorting atomic coordinates in lexicographical order. The TB Hamiltonian has been constructed for a cubic-shaped hydrogen-passivated silicon nanocrystal consisting of 516 atoms.}
    \label{fig:lexi}
\end{figure}

 \begin{figure}[t]
    \centering
    \includegraphics{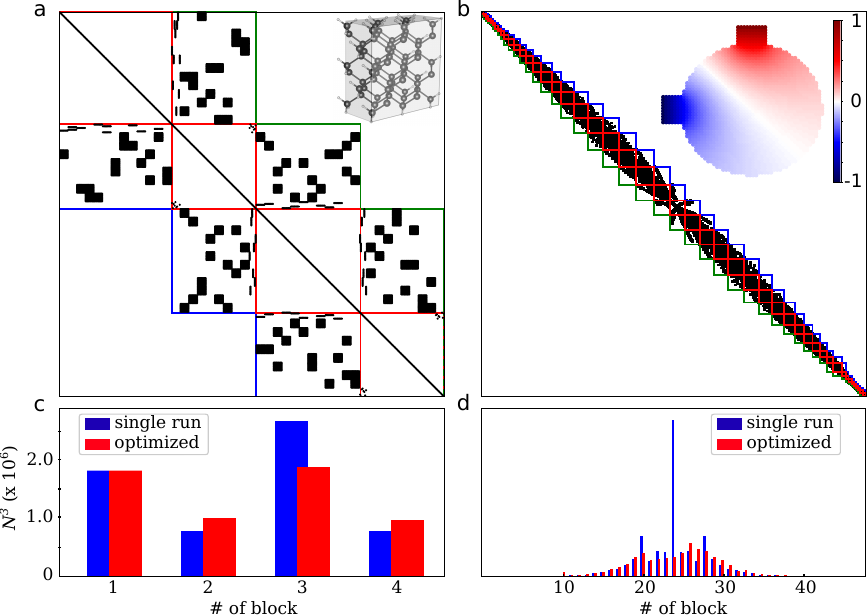}
    \caption{Sparsity patterns and detected block-tridiagonal structure of TB matrices for a) the hydrogen passivated silicon nanowire with 77 atoms in the unit cell and sp$^3$d$^5$s$^*$ basis set for silicon atoms and single s-orbital for hydrogen, and b) 2D quantum billiard consisting of 1888 atoms, each has a single s-orbital respectively. c) and d) distribution of block size cubes for silicon nanowire and quantum billiard respectively computed with a greedy algorithm and optimized recursive algorithm.}
    \label{fig:tri_block}
\end{figure}

The sparsity pattern of the Hamiltonian matrix $H$ depends on the order of sites taken from the list $L$. This statement is illustrated in Fig. \ref{fig:intro}; the order that we store the sites of the system in the Hamiltonian matrix results in different sparsity patterns. Sorting is performed by a function specified by a user during instantiation of the Hamiltonian object. Alternatively, one can use one of the function embedded in the framework. If the sorting function is not specified the order of matrix elements follows the order of entries in the input list of atomic coordinates. In the current version we have implemented three sorting routines: sorting by lexicographical order over atomic coordinates, sorting using projections of the position vectors on an arbitrary vector as sort keys, and sorting over keys determined by a potential function over atomic coordinates.

Sorting by lexicographical order implies that atoms are arranged in a sequence of slices along an axis specified by the first element in the tuple of coordinates. Although this kind of sorting does not guarantee the minimal bandwidth, in most cases it results in matrices with relatively narrow bandwidth (see Fig. \ref{fig:lexi} for an illustrative example).

%3333333333333333333333333333333333333333333333333333333333333333333333333333333333333333333333333%
\begin{algorithm}[!h]
    \SetKwInOut{Input}{Input}
    \SetKwInOut{Output}{Output}
\SetKwComment{com}{}{}
\SetKwFunction{cummax}{cummax}
\SetKwFunction{bandwidth}{bandwidth}
\SetKwFunction{prof}{prof}
\SetKwFunction{max}{max}
\SetKwInput{and}{and}
\SetKwFunction{len}{len}
\SetKwFunction{append}{append}
\SetKwFunction{merge}{merge}
\SetKwFunction{blocks}{compute\_blocks}
\SetSideCommentRight
\DontPrintSemicolon
\caption{Greedy algorithm to compute block-tridiagonal structure of matrix with fixed sizes of leftmost and rightmost blocks \label{a3}}
    function \underline{\texttt{compute\_blocks}} $(l, r, P, P^*, N)$\;
    \Input{Sizes of leftmost and rightmost blocks, $l$ and $r$, profiles of the sparsity pattern, $P$ and $P^*$, and $N$ is the size of $P$.}
    \Output{An array of sizes of diagonal blocks, $blocklist$}
    {
    \tcp{If blocks do not overlap}
    \uIf {$l + r < N$}{
        $new\_l = P[l] - l$ \;
        $new\_r = P^*[r] - r$ \;
        % $new\_r = np.max(np.argwhere(np.abs(P - (N - r)) -
        %                                      np.min(np.abs(P - (N - r))) == 0)) + 1$ \;
        % $new\_r = N - new\_r - r$ \;
        \tcp{Spacing between blocks is sufficient}
        \uIf {($l + new\_l + r \leq N) \And (l + new\_r + r \leq N)$}{

            $blks = \blocks(new\_l, new\_r,$ \\
                            \hspace{3 cm} $P[l+1:N-r] - l,$ \\
                            \hspace{3 cm} $P^*[r+1:N-l] - r,$ \\
                            \hspace{3 cm} $N-r-l)$ \;
                            
            $blocklist \leftarrow [l, blks, r]$ \;
        }
        \tcp{Spacing between blocks is not sufficient}
        \Else{
            \uIf {$new\_l > new\_r$}{
                $blocklist \leftarrow [l, N - l]$ \;
            }
            \Else{
                $blocklist \leftarrow [N - r, r]$ \;
            }
        }
    }
    \tcp{Blocks do overlap}
    \Else{
        $blocklist \leftarrow [N]$ \;
    }
}
\end{algorithm}

%%%%%%%%%%%%%%%%%%%%%%%%%%%%%%%%%%%%%%%%%%%%%%%%%%%%%%%%%%%%%%%%%%%%%%%%%%%%%%%%%%%%%%%

%11111111111111111111111111111111111111111111111111111111111111111111111111111111111111111111111111%    
\begin{algorithm}[!h]
\SetKwInOut{Input}{Input}
\SetKwInOut{Output}{Output}
\SetKwFunction{bandwidth}{bandwidth}
\SetKwFunction{cut}{find\_optimal\_cut}
\SetKwFunction{blocks}{compute\_optimal\_blocks}
\SetSideCommentRight
\DontPrintSemicolon
\caption{Computing optimal block-tridiagonal representation \label{a4}}
function \underline{\texttt{compute\_optimal\_blocks}} $(P, P^*, l, r, N)$\;
\Input{Profiles of the sparsity pattern, $P$ and $P^*$, sizes of leftmost and rightmost blocks, $l$ and $r$, and $N$ is the size of $P$.}
\Output{Array of diagonal blocks,  $blocklist$.}
{
    $j, blocklist, l', r' \leftarrow \cut(P, P^*, l, r, N)$\;
    $flag \leftarrow \textsf{true} $\;
    \tcp{Check if left block is large enough for recursive calls to continue}
    \uIf{$l + r' < j$}{
        $P_{new} = P[1:j]$\;
        \For{$n \in [1, j]$ }{ 
            $P_{new}^*[n] \leftarrow P^*[N-j+n] - N + j$\;
        }
        $blocklist' \leftarrow \blocks(P_{new}, P_{new}^*, l, r', j)$\;
    }
    \uElseIf{$l + r' = j$}{
        $blocklist' = [l, r']$
    }
    \Else{
        $flag \leftarrow \textsf{false} $\
    }
    \tcp{Check if right block is large enough for recursive calls to continue}
    \uIf{$l' + r \le N - j$}{
            
        \For{$n \in [1, N-j]$ }{ 
            $P_{new}[n] \leftarrow P[j+n] - j$\;
        }
        $P_{new}^* = P^*[1:N-j]$\;
        % $blocklist'' \leftarrow \blocks(B[j+1:N] + \{1..N-j\}, P^*[1:N-j], l', r)$\;
        $blocklist'' \leftarrow \blocks(P_{new}, P_{new}^*,$\\
        \hspace{1.5cm} $ l', r, N-j)$\;
    }
    \uElseIf{$l' + r = j$}{
        $blocklist'' = [l', r]$
    }
    \Else{
        $flag \leftarrow \textsf{false} $\
    }
    \tcp{Concatenate two lists if they are compatible with the size constraints.  Otherwise the algorithms outputs the block list computed in the line 2}
    \If{flag}{
        $blocklist \leftarrow blocklist' + blocklist''$
    }
}
\end{algorithm}
%%%%%%%%%%%%%%%%%%%%%%%%%%%%%%%%%%%%%%%%%%%%%%%%%%%%%%%%%%%%%%%%%%%%%%%%%%%%%%%%%%%%%%%%%%%%%%%%%%%

%2222222222222222222222222222222222222222222222222222222222222222222222222222222222222222222222222%
\begin{algorithm}[!h]
\SetKwInOut{Input}{Input}
\SetKwInOut{Output}{Output}
\SetKwFunction{bandwidth}{bandwidth}
\SetKwFunction{cut}{find\_optimal\_cut}
\SetKwFunction{blocks}{compute\_blocks}
\SetSideCommentRight
\DontPrintSemicolon
\caption{Finding the best truncation point giving smallest value of the goal function, Eq. (\ref{eq:gf}) \label{a5}}
function \underline{\texttt{find\_optimal\_cut}} $(P, P^*, l, r, N)$\;
\Input{Profiles of the sparsity pattern, $P$ and $P^*$, sizes of leftmost and rightmost blocks, $l$ and $r$, and $N$ is the size of $P$.}
\Output{Index of the optimal cut, $j$, an array of sizes of diagonal blocks, $blocklist$, and sizes of the diagonal blocks adjacent to the cut, $l'$ and $r'$}
{
$L \leftarrow \text{empty list}$ \;
$E \leftarrow \text{empty list}$ \;
\tcp{Loop over all possible cutting indices}
\For{$j \in [l+1, N-r]$ }{ 
    \tcp{Compute sparsity pattern profiles for the left block}
    $P_{new} = P[1:j]$\;
    \For{$n \in [1, j]$ }{ 
        $P_{new}^*[n] \leftarrow P^*[N-j+n] - N + j$\;
    }
    \tcp{Compute sub-block sizes for the left block}
    $blocklist' \leftarrow \blocks(l, P^*[N-j]-N+j,$\\
    \hspace{1.5 cm} $ P_{new}, P_{new}^*, j)$\;
    \tcp{Compute sparsity pattern profiles for the right block}
    $P_{new} = P^*[1:N-j]$\;
    \For{$n \in [1, N-j]$ }{ 
        $P_{new}^*[n] \leftarrow P[j+n] - j$\;
    }
    \tcp{Compute sub-block sizes for the right block}
    $blocklist'' \leftarrow \blocks(r, P[j]-j, P_{new}, P_{new}^*, N-j)$\;
    \tcp{Concatenate two lists}
    $blocklist \leftarrow blocklist' + blocklist''$\;
    $L.\append(blocklist)$  \;
    \tcp{Evaluate the goal function for each index $j$}
    $E.\append\left(\sum\limits_{n\: \in \: blocklist} n^3 \right)$ \;
}
\tcp{Find the optimal cut}
$s = \argmin\limits_j E[j]$\;
$blocklist = L[s]$ \;

}
\end{algorithm}
%%%%%%%%%%%%%%%%%%%%%%%%%%%%%%%%%%%%%%%%%%%%%%%%%%%%%%%%%%%%%%%%%%%%%%%%%%%%%%%%%%%%%%%%%%%%%%%%%%%
The keys computed as projections on a vector can be used for one-dimenssional structures with collinear leads, when we want to arrange atoms into slices cut along a specific direction in Euclidean space (normally parallel to leads). An example of the matrix sparsity pattern with elements sorted according to this method for a silicon nanowire is shown in Fig. \ref{fig:tri_block}a. The presented TB Hamiltonian is for the hydrogen-passivated silicon nanowire with the primitive cell containing 77 atoms with the sp$^3$d$^5$s$^*$ basis set for silicon atoms and a single s-orbital for hydrogen.

The third embedded sorting method can be used for an arbitrary two-terminal device. For this method, the atomic grid is approximated by a capacitance model \cite{cap_model}, where atoms are represented by a charge nodes with interactions between neighbours modeled by a capacitor. In this case, the sorting keys are determined by a vector $V_j$ which is a solution of the matrix equation $CV=R$, where $C$ is the capacitance matrix and $R$ is a charge distribution. The capacitance matrix can be computed from the adjacency matrix, $A$, of the graph describing connectivity of the nodes:

\begin{equation}
     C_{ij}=    \begin{cases}
      -\sum_{j  \ne i} A_{ij}, & \text{if}\ i=j, \\
      A_{ij}, & \text{otherwise},
    \end{cases}
\end{equation}

The elements of vector $R$ equal minus one for nodes contacting with left lead, plus one for nodes contacting with right lead and zero for all other nodes. The matrix equation has been solved using the LGMRES method \cite{Baker}. We illustrate this sorting procedure by applying it to a two-dimenssional two-terminal quantum billiard consisting of 1888 atoms, each has a single orbital. The resulting sparsity pattern is shown in Fig. \ref{fig:tri_block}b.

Note that all three sorting methods are heuristic approaches, meaning that they do not guarantee the minimal bandwidth of TB matrices, but lead to a significant reduction of the bandwidth comparing to an arbitrary ordering of atomic coordinates. Another way to perform sorting of atomic coordinates efficiently can rely on the graph partitioning technique described in \cite{WIMMER20098548}. This technique will be embedded in the codebase in the next release.

\begin{figure}[t]
    \centering
    \includegraphics{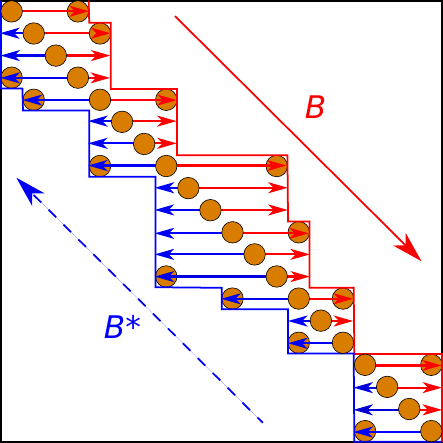}
    \caption{Definition of the matrix bandwidth profiles $B$ and $B^*$ of a matrix $A$.}
    \label{fig:profiles1}
\end{figure}

It is useful for many applications that our matrix be split into a block-tridiagonal structure. In this work we propose a greedy algorithm that can detect optimal block-tridiagonal representation of the band matrix. The proposed algorithm first evaluates the bandwidth of a matrix defined by the expression:
 
 \begin{equation}
    B[i] = \max \left( \left\{j \in K: A_{ij} \neq 0 \right\} \cup \left\{i \right\} \right) - i,
 \end{equation}
where $K$ is the index set of the matrix. The matrix bandwidth is then given by the maximal element of the array $B$. According to this expression the bandwidth is computed for each row as a difference between the largest index of the non-zero elements and the index of the element on the main diagonal (see Fig. \ref{fig:profiles1}). If no such element exists, the bandwidth is equal to zero.  The sparsity pattern profile $P[i]$ is defined by the cumulative maximum of $B[i] + i$. We will also need an auxiliary vector $P^*[i]=B^*[i] + i$ expressed in term of the conjugate bandwidth profile $B^*$ defined as (see Fig. \ref{fig:profiles1}): 
  \begin{equation}
    B^*[i] = i - \min \left(  \left\{j \in K: A_{N-i, j} \neq 0 \right\} \cup \left\{i \right\} \right).   
 \end{equation}

\begin{figure}[t]
    \centering
    \includegraphics{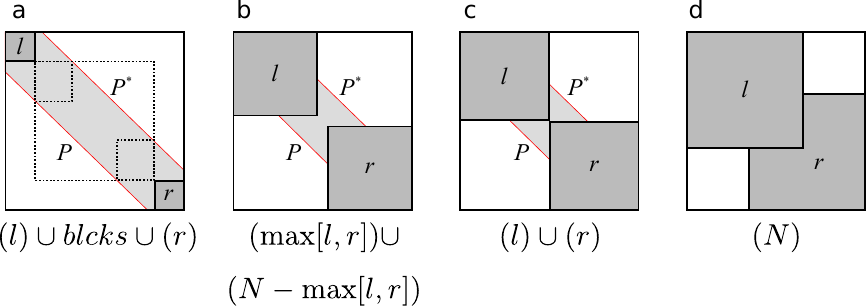}
    \caption{Depending on the size of the leftmost and rightmost blocks, $l$ and $r$, and the size of the matrix $N$, Algorithm \ref{a4} proceeds in one of four possible ways: a) continues recursively if the spacing between blocks is larger than sizes of newly generated blocks, b) return size of the largest block, either $l$ or $r$, and the difference between the size of the matrix and the size of that block if the spacing between blocks is smaller than sizes of newly generated blocks but blocks do not overlap, c) returns $l$ or $r$ if $l+r = N$, and d) returns $N$ if blocks overlap. }
    \label{fig:blocks}
\end{figure}

 Sizes of diagonal blocks of the original matrix are computed using the function \texttt{compute\_blocks($l, r, P, P^*, N$)} which takes fixed sizes of the leftmost and rightmost blocks, $l$ and $r$, as well as the sparsity pattern profiles, $P$ and $P^*$, and the size of the matrix, $N$, as input arguments. The sizes of blocks $l$ and $r$ are defined by the maximum index of non-zero elements in the corresponding coupling Hamiltonians $H_{-1}$ and $H_1$ for the open systems or by $l=P[0]$ and $r=P^*[0]$ for the isolated structures. From the mathematical point of view, these conditions are sufficient for the algorithm to work but they are not necessary, meaning that in some cases they can be relaxed in some cases. Strictly speaking, the size of the coupling matrix should be smaller than the sum of the sizes of the two leftmost or rightmost diagonal blocks. The algorithm \texttt{compute\_blocks} implements Algorithm \ref{a3} according to which if sizes of left and right blocks are small enough (see Algorithm \ref{a3} for explicit conditions), the algorithm defines new smallest leftmost and rightmost blocks that fit into the sparsity pattern profiles. Also at this stage the algorithm determines new sparsity pattern profiles being subsets of the original profiles taken from $l$ to $N-r$, and invokes the function \texttt{compute\_blocks} with new parameters again (see Fig. \ref{fig:blocks}a). At each function call, new block sizes $l$ and $r$ are determined by the sparsity pattern profiles. Since those sizes are the minimal allowed sizes taken at the current step, the algorithm belongs to the class of greedy algorithms \cite{cormen2009introduction}. The sequential calls break in one of three cases illustrated in Fig. \ref{fig:blocks}b-d: the spacing between left and right blocks is not sufficient for the algorithm to continue, the function returns the ordered set of two elements $\left(\max[l, r]\right) \cup \left(N - \max[l, r]\right)$  (see Fig. \ref{fig:blocks}b); the sum of the left and right block sizes is equal to the size of original matrix, the function returns an ordered set consisting of two elements $\left(l, r\right)$ (see Fig. \ref{fig:blocks}c); and the left and right blocks overlap, the function returns the size of the original matrix $\left(N\right)$ (see Fig. \ref{fig:blocks}d).
 
\begin{figure}[!t]
    \centering
    \includegraphics{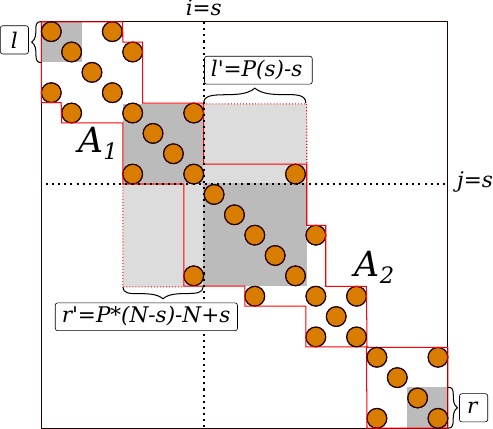}
    \caption{Illustration of a single step of the recursive Algorithm \ref{a4} that computes optimized block-tridiagonal representation of the matrix. At this step matrix $A$ is split in two subblocks $A_1$ and $A_2$; sizes of blocks $r'$ and $l'$ are computed from the sparsity pattern profiles.}
    \label{fig:profiles2}
\end{figure}

A greedy algorithm does not guarantee the optimal solution. Indeed, during analysis of our algorithm we observed instances that lead to very large block sizes when the bandwidth varies substantially over the matrix. However, we found that the solution can be further improved by sequentially splitting the largest blocks. In order to do that we start with splitting the original matrix $A$ into two sub-matrices, $A_1$ and $A_2$ (see Fig. \ref{fig:profiles2}), determined by a value of some cutting index, $s$ (see Algorithm \ref{a4}). At the cutting point, the sparsity pattern profiles determine sizes of new rightmost and leftmost blocks, $r'$ and $l'$, for matrices $A_1$ and $A_2$ shown by grey shaded area in Fig. \ref{fig:profiles2}. To determine the most favorable choice of the cut $s$, we apply the previously discussed algorithm \texttt{compute\_blocks} to each of the matrices $A_1$ and $A_2$ for a range of values $s$ in the interval $(l, N-r)$ and minimize the goal function:
 \begin{equation}
 E(s)= \sum_{j=1}^{s-1} N_j^3+\sum_{j=s}^N N_j^3,    
 \label{eq:gf}
 \end{equation}
 where $N_j$ is the size of $j$-th block, (see Algorithm \ref{a5} implementing the function  \texttt{find\_optimal\_cut}). Minimization of the goal function represented by a sum of cubes of diagonal blocks optimizes the computation time for recursive Green's function algorithms which scale as $\mathcal{O}(N_j^3)$. If the sizes of the matrices $A_1$ and $A_2$ are larger than a sum of sizes of leftmost and rightmost blocks, further improvement can be achieved by repeating this procedure recursively for each of the matrices $A_1$ and $A_2$ (see Algorithm \ref{a4}). We find the points of these optimal cuts are almost always where the bandwidth is at its widest.
 
We have applied the algorithms described above for two structures whose TB matrices have sparsity patterns shown in Fig. \ref{fig:tri_block} a and b. In the first case, the original size of the matrix is $446 \times 446$. The algorithm outputs the diagonal block sizes of 131, 97, 121, and 97 (shown in Fig. \ref{fig:tri_block}a as red rectangles). Note that the algorithm suggests a solution with four blocks, and this corresponds to the number of silicon atomic layers in the primitive cell along the direction of translation. For the second case with the matrix size $1888 \times 1888$, the number of blocks is 51. In Fig. \ref{fig:tri_block} c and d we also compare output of the function \texttt{compute\_blocks} that computes the block sizes in a single run with the output of the function \texttt{compute\_optimal\_blocks} computing an optimized solution. In the last case the sizes of block are more equalized.
 
\section{NanoNET framework}

 \begin{figure*}[t]
    \centering
    \includegraphics{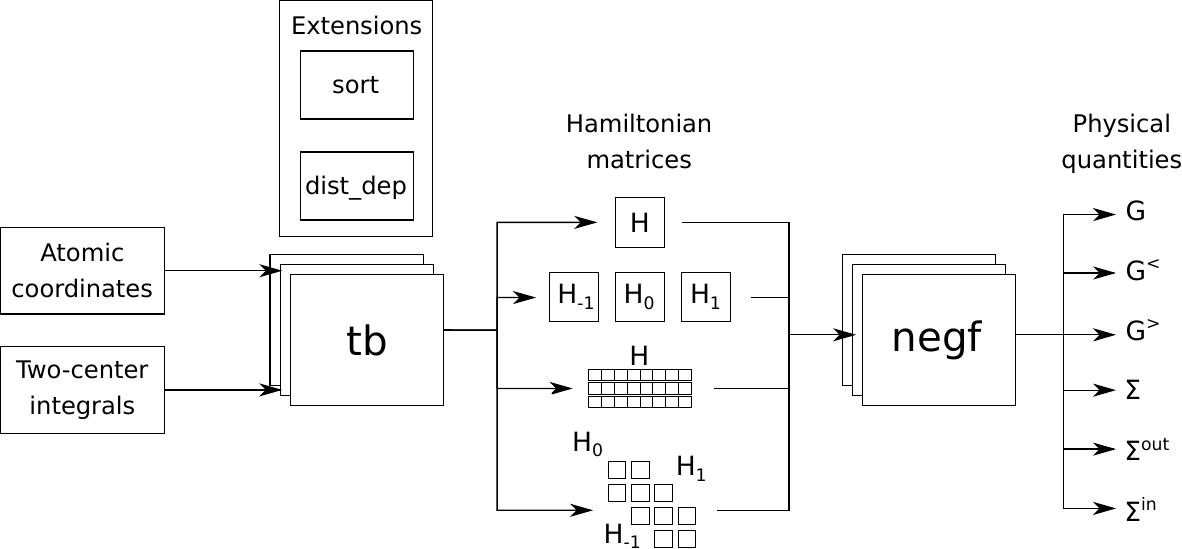}
    \caption{A generic control flow for applications developed with \textit{NanoNET}.}
    \label{fig:nanonet_app}
\end{figure*}

\subsection{Software Architecture}

A generic control flow for applications developed with \textit{NanoNET} \cite{Nanonet} is shown schematically in Fig. \ref{fig:nanonet_app}. The input parameters are the list of atomic coordinates and a table of two-center integrals. The framework contains two packages \texttt{tb} and \texttt{negf}. The package \texttt{tb} is the core responsible for composing and diagonalizing Hamiltonian matrices. The \texttt{negf} package processes TB matrices; it contains subroutines for computing Green's functions, namely implementing the Recursive Green's Function (RGF) algorithm \cite{negf}.

\texttt{tb} represents a library of Python classes facilitating building Hamiltonian matrices, imposing periodic boundary conditions and computing electronic structure for molecules and solids using the TB method in the single-particle approximation. The Hamiltonian matrices are built from an XYZ file containing atomic coordinates and a list of TB parameters. 

\begin{figure}[!h]
    \centering
    \includegraphics{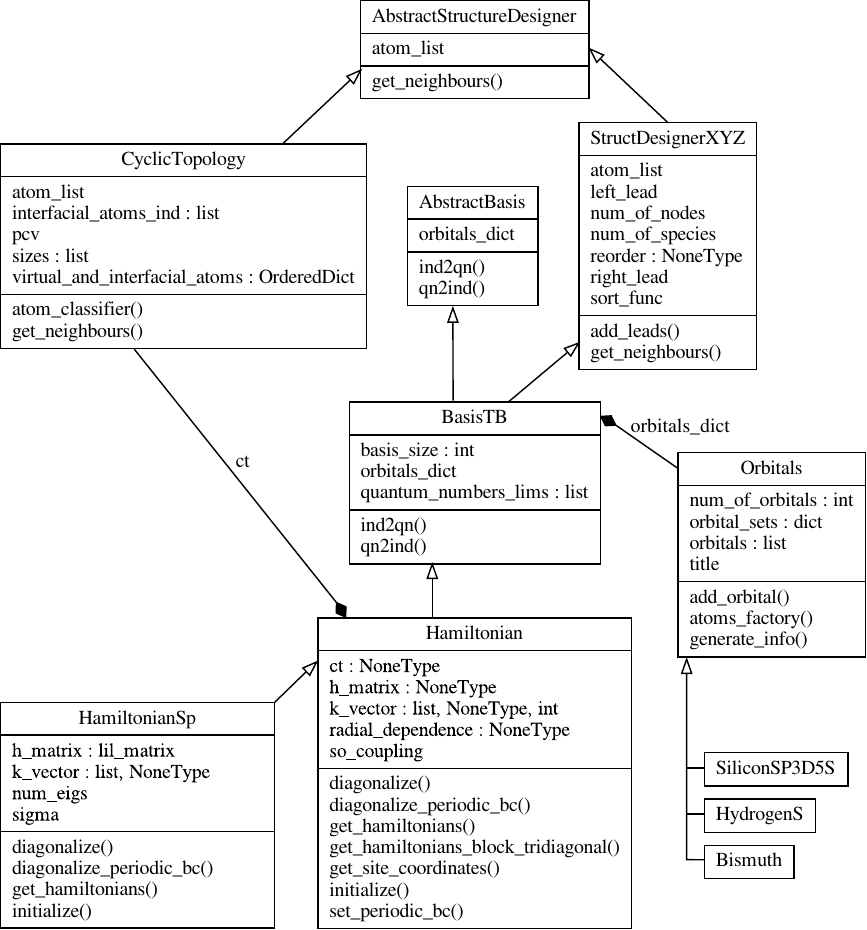}
    \caption{UML diagram of the subpackage \texttt{tb} of \textit{NanoNET}}
    \label{fig:uml}
\end{figure}

The software architecture relies on the object-oriented paradigms --- the framework represents a library of classes whose UML diagram is shown in Fig. \ref{fig:uml}. The central class of the framework is called \texttt{Hamiltonian} and contains all necessary information and functionality to construct Hamiltonian matrices that represents its main goal. This class inherits properties and member functions from classes \texttt{BasisTB} and \texttt{StructureDesignerXYZ} --- abstractions for basis sets and geometrical configuration of atoms correspondingly. The first one stores a look-up table that allows one to associate a set of orbitals with a label denoting a chemical element. The second one stores a $kd$-tree built from the list of atomic coordinates.

The class \texttt{CyclicTopology} is used when periodic boundary conditions are applied. It translates atomic coordinates according to translation vectors and creates a $kd$-tree for atoms outside the primitive cell, needed to compute the Hamiltonian matrix responsible for coupling between neighbouring primitive cells.

The orbital sets are created using facilities of the class \texttt{Orbitals}. This class is the parent class for all basis sets. The current version of the software package contains predefined basis sets: sp$^3$d$^5$s$^*$ model for silicon \cite{Jancu}, \texttt{SiliconSP3D5S}, single s-orbital for hydrogen \cite{Zheng}, \texttt{HydrogenS}, and an sp$^3$ model for bismuth \cite{Liu}, \texttt{Bismuth}. 

The version of the class \texttt{Hamiltonian} that uses sparse matrix representations is implemented in sub-class \texttt{HamiltonianSp} having the same interface with some redefined member-functions.

There are also supplementary modules in \texttt{tb} such as \texttt{diatomic\_matrix\_element} and \texttt{block\_tridiagonalization} which are used in the class \texttt{Hamiltonian} and contain a set of procedures for computing the TB matrix elements and block-tridiagonal structure of a matrix respectively.

The package \texttt{negf} is written in the procedural programming paradigm and contains functions that computes complex band structure, self-energies of leads and non-equilibrium Green's functions.

\subsection{Software Functionalities}

The interface to \textit{NanoNET} is provided either by an application programming interface represented by a library of Python classes, or by a command line user interface for Unix-like operating systems.

The functionality of the package includes:
\begin{itemize}[noitemsep, nolistsep]
    \item building TB Hamiltonian matrices (Algorithms \ref{a1} and \ref{a2});
    \item detecting block-tridiagonal matrix structure (Algorithms \ref{a3} -- \ref{a5});
    \item diagonalizing Hamiltonian matrices using standard numerical methods implemented in \emph{NumPy} and \emph{SciPy} packages;
    \item computing a set of non-equilibrium Green's functions as well as observables that can be derived from them such as density of states, transmission function and charge density. The Green's functions are computed using the recursive Green's function algorithm \cite{negf}.
\end{itemize}

\section{Extracting physical quantities of interest from the Hamiltonian matrix and Green's functions}

Although \textit{NanoNET} can be interfaced with any other software by extracting Hamiltonian matrices in any favorable format and passing them as an input to other programs, for the sake of completeness we provide basic routines for computing some physical quantities of interest from Hamiltonian matrices. Time-independent observables, such as energy spectrum, density of states, and stationary and/or equilibrium electron density distribution, can be extracted from the eigenvalues and eigenvectors of the Hamiltonian matrix obtained by the direct matrix diagonalization or computed via the Green's functions formalism that requires computing the matrix inverse. The direct matrix diagonalizaion is implemented as a member function of the class \texttt{Hamiltonian} in the package \texttt{tb}, while all other algorithms performing post-processing of the tight-binding matrices are placed in the package  \texttt{negf}.

\subsection{Band structure}

For the periodic structures the band structure can be found solving the eigenvalue problem $H(\mathbf{k}) \psi_{\mathbf{k}} = E(\mathbf{k})\psi_{\mathbf{k}}$, where $\psi_{\mathbf{k}}$ and $E(\mathbf{k})$ are the eigenvectors and eigenvalues of the TB matrix, and the wave vector $\mathbf{k}$ is set as a parameter. This problem is solved by a direct matrix diagonalization performed in the class \texttt{Hamiltonian} using LAPACK \cite{laug} routines with a Python interface provided via \emph{Numpy} package.

\subsection{Complex band structure}

Recently much attention has been drawn to the complex band structure computation in the context of molecular junction transmission \cite{Jensen} and Majorana modes of nanowires \cite{Osca2019}. The complex band structure can be computed from the TB Hamiltonian written in the form: $H(\mathbf{k}) = H_0 + H_{-1}e^{-ik} + H_1e^{ik}$, where $H_0$ is the intracell TB Hamiltonian and $H_{-1}$ and $H_1$ are the  Hamitonians of intercell coupling. This leads to the eigenvalue problem: 
$$\left[ E - H_0 - H_{-1}e^{-ik} - H_1e^{ik}  \right] \psi_E = 0.$$
Here, instead of the wave vector, we can set energy $E$ as a parameter and find the eigenvalues in the form $\lambda = e^{ik}$ \cite{PhysRevB.25.3975, Wimmer} from the matrix:

\begin{equation}
\begin{bmatrix} 
0 & 1 \\
-H_1^{-1}H_{-1} & H_1 \left(E-H_0 \right) 
\end{bmatrix}
\begin{bmatrix} 
\psi_E\\
\lambda_E \psi_E
\end{bmatrix} = \lambda_E
\begin{bmatrix} 
\psi_E\\
\lambda_E \psi_E.
\end{bmatrix}
\label{eq:cbs}
\end{equation}

\subsection{Self-energies for periodic leads}

A quasi periodic structure is often used to model electrodes or leads that can be attached to a nanostructure. The leads modify the nanostructure by providing an additional source of elastic and/or inelastic scattering. The effect of the leads on the Hamiltonian or Green's functions associated with the nanostructure can be conveniently expressed through the self-energies. In the case of the two-terminal devices we have left and right leads whose self-energies can be expressed in terms of the eigenvectors and eigenvalues obtained from Eq. (\ref{eq:cbs}). Following the procedure described in \cite{Wimmer}, the eigenvectors $\psi_E$ and eigenvalues $\lambda_E$ can be divided in two classes, left- and right propagating modes, depending on the position of complex eigenvalues in the complex plane. The eigenvectors and eigenvalues for the left-propagating modes are collected in matrices $\Psi_<$ and $\Lambda_<$ respectively, while the  right-propagating modes are described by the matrices $\Psi_>$ and $\Lambda_>$. With this notation, the self-energies describing couplings to the semi-infinite leads reads \cite{Wimmer}:

\begin{equation}
	\Sigma_L = H_1 \Psi_> \Lambda_> \Psi_>^{-1}, \hspace{1cm} \Sigma_R = H_{-1} \Psi_< \Lambda_> \Psi_<^{-1}.
\end{equation}

Another set of useful quantities employed in the non-equilibrium Green's function computations are the in-scattering and out-scattering matrices:

\begin{equation}
\Sigma_j^\text{in} = \Gamma_j f_j(E, \mu_j), \hspace{0.7cm} \Sigma_j^\text{out} = \Gamma_j \left[1-f_j(E, \mu_j)\right],
\end{equation}
where $j \in \{L, R\}$, and $f_j(E, \mu_j)$ is the Fermi-Dirac distribution function with the chemical potential $\mu_j$, and $\Gamma_j = i\left(\Sigma_j - \Sigma_j^{\dagger} \right)$ is the coupling matrix.

\subsection{Green's functions}

Computing Green's functions is based on the recursive Green's function algorithm published in \cite{negf}. The algorithm implies a steady-state regime of the electron transport and discretized real space. The program outputs following functions as a result: retarded Green's function $G = \left[EI - H - \Sigma \right]^{-1}$, electron correlation function $G^n = G \Sigma^\text{in} G^{\dagger}$ and hole correlation function $G^p = G \Sigma^\text{out} G^{\dagger}$. In the literature \cite{Stefanucci, Bonitz}, the $G^n$ and $G^p$ are also known as the lesser and greater Green's functions, $-i G^<$ and $+i G^>$, respectively.

The retarded Green's function contains information related to properties of a system in thermodynamic equilibrium such as density of states,
\begin{equation}
\mathrm{DOS}(\varepsilon) = \frac{i}{\pi} \mathrm{Tr} \left( G-G^{\dagger}\right),
\end{equation}
transmission function (Caroli formula \cite{Klymenko}),
\begin{equation}
T(\varepsilon) =\mathrm{Tr} \left[ G^{\dagger} \Gamma_R G  \Gamma_L  \right],
\end{equation}
and conductance in the linear transport regime \cite{Landauer},

\begin{equation}
	g(\varepsilon) = -\frac{2e^2}{\hbar} \int d\varepsilon  \left. \frac{d f_0(\varepsilon)}{d\varepsilon}  \right|_{\varepsilon_F} T(\varepsilon).
	\label{eq:landauer}
\end{equation}
where $f_0(\varepsilon)$ is the equilibrium distribution function, and the pre-factor 2 corresponds to spin-degeneracy.

The electron and hole correlation functions can be used to compute physical parameters such as current density, associated with the non-equilibrium regime \cite{negf}.

It is necessary to note that most of the mentioned above transport properties can be also computed using the spectral methods \cite{Zheng, Neophytou2008, Lansbergen2008, PhysRevB.73.165319}. Such methods do not require matrix inversion. The efficiency of the spectral methods can be significantly enhanced using the Chebyshev expansion method \cite{FAN201428, KITE}, kernel polynomial method \cite{KPM, TBTK, pybinding, kwant} or reduced mode space techniques \cite{Afzalian_2018, PhysRevB.85.035317} which is planned for a future release.

\section{NanoNET: illustrative examples}

As an example of \textit{NanoNET} usage we have computed the electronic structure for a silicon nanowire, discussed in Section 2.5, and bulk bismuth. The silicon nanowire represents the case of 1D periodic structure with large number of atoms in the primitive cell (the atomic coordinates of atoms in the primitive cell are shown in Supplementary material).  The bulk bismuth is a representative example where second and third nearest neighbour interactions have to be taken into account.  

\subsection{Silicon nanowire example}

First we load all necessary modules.

\begin{python}
import numpy as np
import nanonet.tb as tb
import nanonet.negf as negf
\end{python}

Then, we set the primitive cell parameters and inquire usage of predefined basis sets for silicon and hydrogen.

\StartLineAt{4}
\begin{python}
tb.Orbitals.orbital_sets = {'Si': 'SiliconSP3D5S',
                            'H': 'HydrogenS'}
\end{python}

In this example we will use a custom sorting procedure for atomic coordinates defined below:

\StartLineAt{6}
\begin{python}
def sorting(coords, **kwargs):
    return np.argsort(coords[:, 2], 
                  kind='mergesort')
\end{python}

Here, we make an instance of the class \texttt{Hamiltonian} via reading atomic coordinates from the XYZ file.

\StartLineAt{9}
\begin{python}
h = tb.Hamiltonian(xyz='SiNW2.xyz', 
                   nn_distance=2.4,
                   sort_func=sorting).initialize()
\end{python}

The content of the file 'SiNW2.xyz' is shown in the Appendix. The object constructor also takes the argument \texttt{nn\_distance} which is the maximal distance determining the nearest neighbours, measured in Angstroms. The actual computations of the Hamiltonian matrix elements are performed by invoking the member function \texttt{initialize()} of an object of the class \texttt{Hamiltonian}. After initialization of the Hamiltonian matrix, the periodic boundary conditions are applied:

\StartLineAt{12}
\begin{python}
a_si = 5.50
primitive_cell = [[0, 0, a_si]]
h.set_periodic_bc(primitive_cell)
\end{python}

At this point we may verify the correctness of computed matrices. To do that, we reproduce the band structure of a silicon nanowire published previously in \cite{Zheng}.

\StartLineAt{15}
\begin{python}
num_points = 20
k_array = np.linspace(0, np.pi / a_si, num_points)
band_sructure = []

for kk in k_array:
    e, _ = h.diagonalize_periodic_bc([0, 0, kk])
    band_sructure.append(e)
\end{python}

\begin{figure}[t]
    \centering
    \includegraphics{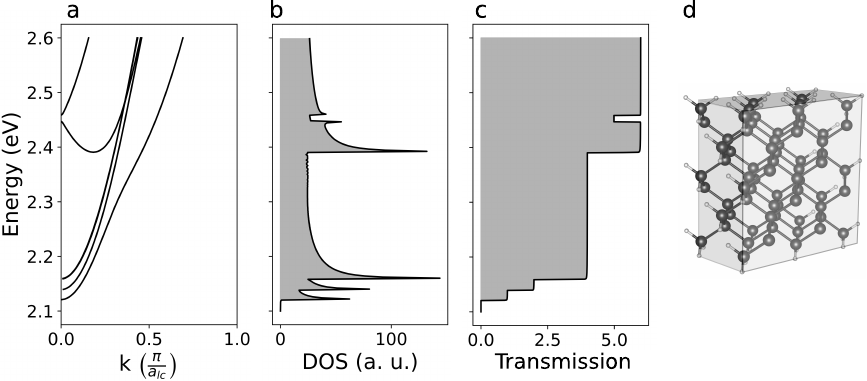}
    \caption{a) Band structure, b) density of states and c) electron transmission function of the hydrogen passivated silicon nanowire using the semi-empirical tight-binding method  implemented in \textit{NanoNET} framework. The primitive cell of the nanowire consisting of 77 atoms is shown in the panel d.}
    \label{fig:transport}
\end{figure}

The resulting band structure is visualized in Fig. \ref{fig:transport}a. Now we proceed to computing the electron transmission function. 

In order to get the Hamiltonian matrices $H_0$, $H_{-1}$ and $H_{1}$ for Eq. (\ref{eq:cbs}) and evaluate their block-tridiagonal structure, we invoke the member function \texttt{get\_hamiltonians\_block\_tridiagonal()}:

\StartLineAt{22}
\begin{python}
hl_b, h0_b, hr_b,\
blocks = h.get_hamiltonians_block_tridiagonal()
\end{python}

Now we make an energy array and initialize arrays for the density of states and transmission spectrum.

\StartLineAt{24}
\begin{python}
energy = np.linspace(2.1, 2.6, 100)
tr = np.zeros((energy.shape[0]))
dos = np.zeros((energy.shape[0]))
\end{python}

Next, in the energy loop we compute self-energies of leads, $L$ and $R$, their spectral functions, density of states and transmission functions defined by the Caroli formula \cite{Klymenko}.

\StartLineAt{27}
\begin{python}
damp = 0.001j
for j, E in enumerate(energy):
    L, R = negf.surface_greens_function(E,
                                        hl_bd,
                                        h0_bd, 
                                        hr_bd, 
                                        iterate=True, 
                                        damp=damp)
    g_trans, grd, grl, gru,\
    gr_left = negf.recursive_gf(E,
                                hl_bd,
                                h0_bd,
                                hr_bd,
                                left_se=L,
                                right_se=R,
                                damp=damp)
    blks = len(grd)
    for jj in range(blks):
        dos[j] -= np.trace(np.imag(grd[jj])) / blks
    
    gamma_l = 1j * (L - L.conj().T)
    gamma_r = 1j * (R - R.conj().T)
    
    tr[j] = np.real(np.trace(gamma_l @\
                             g_trans @\
                             gamma_r @\
                             g_trans.conj().T)))
\end{python}

The resulting density of states and transmission function are visualized in Fig. \ref{fig:transport}b and c respectively.

\subsection{Bulk Bi crystal example}

\StartLineAt{1}
\begin{python}
from nanonet.tb import Hamiltonian, Orbitals
from nanonet.tb import set_tb_params, get_k_coords
\end{python}

First, we form $sp^3$ LCAO basis set for bismuth atoms \cite{Liu}. The basis set is represented by an instance of the class \texttt{Orbitals}. Each orbital possesses a label, energy value, and a set of quantum numbers. Note, the basis set includes orbitals with different spins since the spin-orbit coupling has to be taken into account.

\StartLineAt{3}
\begin{python}
bi_orb = Orbitals('Bi')
bi_orb.add_orbital("s", energy=-10.906,
                   principal=0, orbital=0,
                   magnetic=0, spin=0)
bi_orb.add_orbital("px", energy=-0.486,
                   principal=0, orbital=1,
                   magnetic=-1, spin=0)
bi_orb.add_orbital("py", energy=-0.486,
                   principal=0, orbital=1,
                   magnetic=1, spin=0)
bi_orb.add_orbital("pz", energy=-0.486,
                   principal=0, orbital=1,
                   magnetic=0, spin=0)
bi_orb.add_orbital("s", energy=-10.906,
                   principal=0, orbital=0,
                   magnetic=0, spin=1)
bi_orb.add_orbital("px", energy=-0.486,
                   principal=0, orbital=1,
                   magnetic=-1, spin=1)
bi_orb.add_orbital("py", energy=-0.486,
                   principal=0, orbital=1,
                   magnetic=1, spin=1)
bi_orb.add_orbital("pz", energy=-0.486,
                   principal=0, orbital=1,
                   magnetic=0, spin=1)
\end{python}

The atomic coordinates may be defined in a XYZ file or directly in a Python variable containing a formated text string like in the example below.

\StartLineAt{29}
\begin{python}
xyz_coords = """2
Bi2 cell
Bi1       0.0    0.0    0.0
Bi2       0.0    0.0    5.52321494
"""
\end{python}

At this stage all is set to define the instance of the class Hamiltonian:

\StartLineAt{34}
\begin{python}
h = Hamiltonian(xyz=xyz_coords,
                nn_distance=4.6, so_coupling=1.5)
\end{python}

Here the variable \texttt{so\_coupling} specifies the value of the spin-orbit coupling.

For the bismuth crystals, coupling between first, second and third nearest neighbours have to be taken into account. For that we need to specify a so-called radial-dependence function that classifies nearest neighbours in one of mentioned above categories depending on the inter-nuclei distance.

\StartLineAt{36}
\begin{python}
import numpy as np
def radial_dep(coords):

    norm_of_coords = np.linalg.norm(coords)
    if norm_of_coords < 3.3:
        return 1
    elif 3.7 > norm_of_coords > 3.3:
        return 2
    elif 5.0 > norm_of_coords > 3.7:
        return 3
    else:
        return None
\end{python}

The function returns one of four labels. The TB parameters can be set via the function \texttt{set\_tb\_params}. This function follows certain naming convenction for the TB parameters: \texttt{PARAMS\_<el1>\_<el2><order>}, where \texttt{<el1>} and \texttt{<el2>} are chemical elements of a pair of atoms, and \texttt{<order>} is a number specifying the order of neighbours.

The TB parameters and orbital energies for Bi are taken from \cite{Liu}:

\StartLineAt{48}
\begin{python}
# 1NN - Bi-Bi
PAR1 = {'ss_sigma': -0.608,
        'sp_sigma': 1.320,
        'pp_sigma': 1.854,
        'pp_pi': -0.600}

# 2NN - Bi-Bi
PAR2 = {'ss_sigma': -0.384,
        'sp_sigma': 0.433,
        'pp_sigma': 1.396,
        'pp_pi': -0.344}

# 3NN - Bi-Bi
PAR3 = {'ss_sigma': 0,
        'sp_sigma': 0,
        'pp_sigma': 0.156,
        'pp_pi': 0}
set_tb_params(PARAMS_BI_BI1=PAR1, 
              PARAMS_BI_BI2=PAR2,
              PARAMS_BI_BI3=PAR3)
\end{python}

The function \texttt{initialize()} takes a radial-dependence function as an argument. If it is not specified the \texttt{None} label is used for the TB parameters.

\StartLineAt{68}
\begin{python}
h.initialize(radial_dep)
\end{python}

The computed Hamiltonian matrix is related to the isolated nanocrystal whose atomic coordinates are stored in the variable \texttt{xyz\_coords}. The crystal structure is defined setting the periodic boundary conditions with translation vectors defining a primitive cell.

\StartLineAt{69}
\begin{python}
primitive_cell = [[-2.267, -1.309,  3.932],
                  [ 2.267, -1.309,  3.932],
                  [ 0.000,  2.617,  3.932]]
                  
h.set_periodic_bc(primitive_cell)
\end{python}

The band structure is computed for a set of points in the k-space defined below (see Ref. \cite{PhysRev.137.A871} for a definition of these high symmetry points):

\StartLineAt{74}
\begin{python}
sym_k_points = {'GAMMA':  [0.00,  0.00,  0.00],
                'K':      [0.36, -0.80,  0.53],
                'L':      [0.69, -0.40,  0.27],
                'LAMBDA': [0.00,  0.00,  0.40],
                'T':      [0.00,  0.00,  0.80],
                'U':      [0.54, -0.31,  0.80],
                'W':      [0.36, -0.62,  0.80],
                'X':      [0.00, -0.80,  0.53]}
path = ['K', 'GAMMA', 'T', 'W', 'L', 'LAMBDA']
sizes = [10, 10, 10, 10, 10]                
k_array = get_k_coords(path, sizes, sym_k_points)
\end{python}

A computation of the band structure is performed by diagonalizing Hamiltonian matrices for each k-point.

\StartLineAt{85}
\begin{python}
band_structure = []
for k_point in k_array:
    [e, _] = h.diagonalize_periodic_bc(k_point)
    band_structure.append(e)
\end{python}

\begin{figure}
    \centering
    \includegraphics{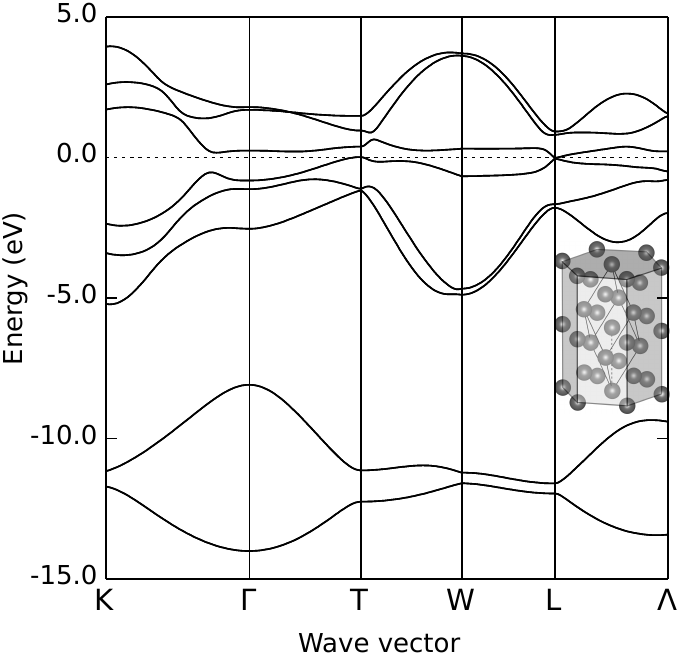}
    \caption{Band structure of bulk bismuth computed using the semi-empirical tight-binding method  implemented in \textit{NanoNET} framework.}
    \label{fig:bs}
\end{figure}

The resulted band structure is shown in Fig. \ref{fig:bs}.

\section{Software performance}
The overall execution time of the applications built with the \textit{NanoNET} framework depends on the considered atomic structure and choice of algorithms. To illustrate this point, we measure the execution time of the components of the script computing the transmission probability for an electron in the Si nanowire from the Section 5.1 with different geometrical sizes. All numerical experiments have been carried out on one processor core of a 24-core Intel Xeon Scalable `Cascade Lake` processor.

\begin{figure}[!h]
    \centering
    \includegraphics{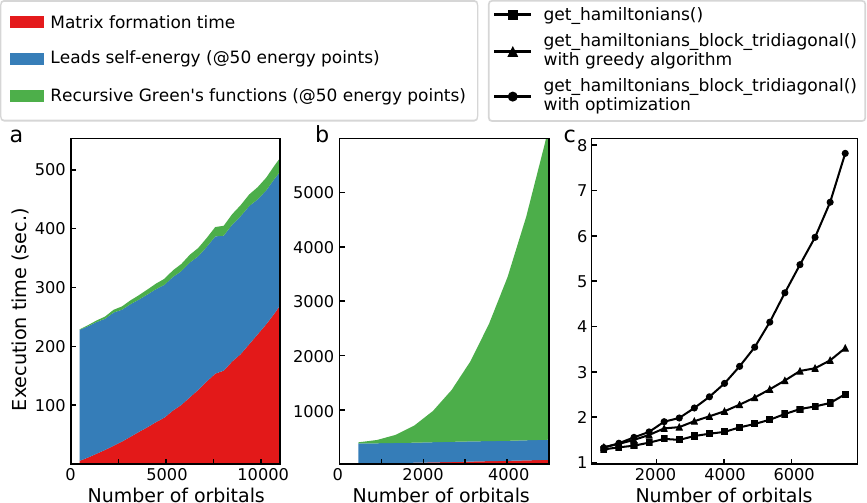}
    \caption{Stack diagrams for the execution time of computing transport properties of the silicon nanowire a) using algorithm detecting tri-blockdiagonal structure and b) without detecting tri-blockdiagonal structure. c) The execution time spent detecting the tri-blockdiagonal structure of the Hamiltonian matrices.}
    \label{fig:times}
\end{figure}

The time required to form the Hamiltonian matrices is determined by the nearest-neighbour search based on the kd-tree. Building such tree scales as $ N\log N$ with the number of orbitals (see red shaded area in  see Fig. \ref{fig:times} a and b). The most favorable way to increase the number of atoms in this structure from the computational point of view is increasing the length of the nanowire in the direction perpendicular to the leads. Since the area of the leads does not change in this case, the computational time spent for the leads self-energy is independent of the number of orbitals (see blue shaded area). The execution time for the recursive Green's function algorithm in this case depends linearly on the number of orbitals, since the algorithm for detecting the tri-blockdiagonal structure of the Hamiltonian matrix finds new blocks when more orbitals are added (see Fig. \ref{fig:times} a). In the limit of extremely long nanowires the time spent for the recursive Green's function algorithm can overcome the time needed to compute the self-energies. Note that building Hamiltonian matrices has to be done only once at the beginning of the script while the self-energies and Green's functions have to be computed for each value of quasi-particle energy (the wall time has been evaluated for 50 points). In Fig. \ref{fig:times} b, we show the execution time for the same structure without splitting the Hamiltonian into blocks. In this case, the execution time for the recursive Green's function algorithm grows as $N^3$ becoming dominant for the structures with more than 2000 orbitals. This time is determined by the execution time of LAPACK \texttt{gelsy} algorithm outputting the least-squares solution to a linear matrix equation.

The least favorable case from a computational point of view is when new atoms are added to the leads and to the nanowire in the direction parallel to leads. The algorithm for the computing leads self-energies is based on solving the eigenvalue problem which scales as $N^3$ and is performed by the LAPACK algorithm \texttt{cggev} for the non-symmetric generalized eigenvalue problem \cite{laug}. The recursive Green's functions algorithm also cannot perform efficiently in this case, since adding new atoms leads to growing block sizes.

The data in Fig. \ref{fig:times} c show performance of the algorithms generating matrices in the tri-blockdiagonal form for a set of longitudinal sizes of the nanowire. If none of those algorithms are applied, some time is spent to split the Hamiltonian into the device region part, and left-lead and right-lead coupling parts (the line with square markers). Computing the tri-blockdiagonal structure with the greedy algorithm takes approximately twice as long with the same scaling (the line with triangle markers). The optimization algorithm is aimed to even out the block sizes and it is polynomial in time (the line with circle markers). However, since all the algorithms mentioned above operate on block sizes and the profiles of the sparsity patterns, not the Hamiltonian matrix itself, even in the last case the overall execution time is only a few seconds. When the Hamiltonian matrices are computed the most time is spent on the nearest-neighbour search.  

\section{Conclusions}

We have introduced a novel open-source \emph{Python} framework \textit{NanoNET} for the electronic structure and electron transport modelling based on the tight-binding method. The framework provides user with facilities to build the tight-binding Hamiltonian matrices in dense, sparse or block-tridiagonal forms taking a list of atomic coordinates and a table of two-center integrals as an input. The framework can be extended by a user in several ways: adding subroutines for sorting atomic coordinates in a way that is required by a certain application and defining a radial distance dependence of tight-binding parameters.  \textit{NanoNET} consists of two \emph{Python} packages \texttt{tb} and \texttt{negf}: the first one is for building TB matrices and the latter contains a set of subroutines that compute NEGF and related physical quantities. In future we plan to extend the package \texttt{tb} by additionally facilities to generate reduced models using the tight-binding mode-space technique \cite{PhysRevB.85.035317}.

\section*{Acknowledgments}
The authors acknowledge support of the Australian Research Council through grant CE170100026 and CE170100039. This research was undertaken with the assistance of resources and services from the National Computational Infrastructure, which is supported by the Australian Government.

\bibliography{bib}

\begin{thebibliography}{10}
\expandafter\ifx\csname url\endcsname\relax
  \def\url#1{\texttt{#1}}\fi
\expandafter\ifx\csname urlprefix\endcsname\relax\def\urlprefix{URL }\fi
\expandafter\ifx\csname href\endcsname\relax
  \def\href#1#2{#2} \def\path#1{#1}\fi

\bibitem{Neophytou2008}
N.~Neophytou, A.~Paul, M.~S. Lundstrom, G.~Klimeck, Simulations of nanowire
  transistors: atomistic vs. effective mass models, Journal of Computational
  Electronics 7~(3) (2008) 363--366.

\bibitem{Lansbergen2008}
G.~P. Lansbergen, R.~Rahman, C.~J. Wellard, I.~Woo, J.~Caro, N.~Collaert,
  S.~Biesemans, G.~Klimeck, L.~C.~L. Hollenberg, S.~Rogge, Gate-induced
  quantum-confinement transition of a single dopant atom in a silicon finfet,
  Nature Physics 4 (2008) 656.

\bibitem{PhysRevB.73.165319}
Y.~M. Niquet, A.~Lherbier, N.~H. Quang, M.~V. Fern\'andez-Serra, X.~Blase,
  C.~Delerue, Electronic structure of semiconductor nanowires, Phys. Rev. B 73
  (2006) 165319.

\bibitem{Zheng}
{Yun Zheng}, C.~{Rivas}, R.~{Lake}, K.~{Alam}, T.~B. {Boykin}, G.~{Klimeck},
  Electronic properties of silicon nanowires, IEEE Transactions on Electron
  Devices 52~(6) (2005) 1097--1103.

\bibitem{Afzalian_2018}
A.~Afzalian, T.~Vasen, P.~Ramvall, T.-M. Shen, J.~Wu, M.~Passlack, Physics and
  performances of {III}{\textendash}v nanowire broken-gap heterojunction
  {TFETs} using an efficient tight-binding mode-space {NEGF} model enabling
  million-atom nanowire simulations, Journal of Physics: Condensed Matter
  30~(25) (2018) 254002.

\bibitem{Ryu}
H.~Ryu, S.~Lee, M.~Fuechsle, J.~A. Miwa, S.~Mahapatra, L.~C.~L. Hollenberg,
  M.~Y. Simmons, G.~Klimeck, A tight-binding study of single-atom transistors,
  Small 11~(3) (2015) 374--381.

\bibitem{Smith2015}
J.~S. Smith, D.~W. Drumm, A.~Budi, J.~A. Vaitkus, J.~H. Cole, S.~P. Russo,
  Electronic transport in si:p $\ensuremath{\delta}$-doped wires, Phys. Rev. B
  92 (2015) 235420.

\bibitem{Rudenko}
A.~N. Rudenko, S.~Yuan, M.~I. Katsnelson, Toward a realistic description of
  multilayer black phosphorus: From $gw$ approximation to large-scale
  tight-binding simulations, Phys. Rev. B 92 (2015) 085419.

\bibitem{PhysRevB.58.7260}
M.~Elstner, D.~Porezag, G.~Jungnickel, J.~Elsner, M.~Haugk, T.~Frauenheim,
  S.~Suhai, G.~Seifert, Self-consistent-charge density-functional tight-binding
  method for simulations of complex materials properties, Phys. Rev. B 58
  (1998) 7260--7268.

\bibitem{Elstner}
M.~Elstner, T.~Frauenheim, E.~Kaxiras, G.~Seifert, S.~Suhai, A self-consistent
  charge density-functional based tight-binding scheme for large biomolecules,
  physica status solidi (b) 217~(1) (2000) 357--376.

\bibitem{Calogero_2018}
G.~Calogero, N.~R. Papior, P.~B{\o}ggild, M.~Brandbyge, Large-scale
  tight-binding simulations of quantum transport in ballistic graphene, Journal
  of Physics: Condensed Matter 30~(36) (2018) 364001.

\bibitem{Bowler_1997}
D.~R. Bowler, M.~Aoki, C.~M. Goringe, A.~P. Horsfield, D.~G. Pettifor, A
  comparison of linear scaling tight-binding methods, Modelling and Simulation
  in Materials Science and Engineering 5~(3) (1997) 199--222.

\bibitem{PhysRevB.51.10157}
E.~Hern\'andez, M.~J. Gillan, Self-consistent first-principles technique with
  linear scaling, Phys. Rev. B 51 (1995) 10157--10160.

\bibitem{PhysRevB.53.12733}
A.~F. Voter, J.~D. Kress, R.~N. Silver, Linear-scaling tight binding from a
  truncated-moment approach, Phys. Rev. B 53 (1996) 12733--12741.

\bibitem{Bonitz}
K.~Balzer, M.~Bonitz, Nonequilibrium Green's Functions Approach to
  Inhomogeneous Systems, Lecture Notes in Physics, Springer Berlin Heidelberg,
  2012.

\bibitem{Stefanucci}
G.~Stefanucci, R.~van Leeuwen, Nonequilibrium Many-Body Theory of Quantum
  Systems: A Modern Introduction, Cambridge University Press, 2013.

\bibitem{PhysRevB.97.085149}
J.~A. Vaitkus, J.~H. Cole, B\"uttiker probes and the recursive green's
  function: An efficient approach to include dissipation in general
  configurations, Phys. Rev. B 97 (2018) 085149.

\bibitem{Datta}
S.~Datta, C.~U. Press, Quantum Transport: Atom to Transistor, Cambridge
  University Press, 2005.

\bibitem{negf}
M.~P. Anantram, M.~S. Lundstrom, D.~E. Nikonov, Modeling of nanoscale devices,
  Proceedings of the IEEE 96~(9) (2008) 1511--1550.

\bibitem{Ozaki}
T.~Ozaki, Efficient low-order scaling method for large-scale electronic
  structure calculations with localized basis functions, Phys. Rev. B 82 (2010)
  075131.

\bibitem{NEMO}
S.~{Steiger}, M.~{Povolotskyi}, H.~{Park}, T.~{Kubis}, G.~{Klimeck}, Nemo5: A
  parallel multiscale nanoelectronics modeling tool, IEEE Transactions on
  Nanotechnology 10~(6) (2011) 1464--1474.

\bibitem{kwant}
C.~W. Groth, M.~Wimmer, A.~R. Akhmerov, X.~Waintal, Kwant: a software package
  for quantum transport, New Journal of Physics 16~(6) (2014) 063065.

\bibitem{pybinding}
D.~Moldovan, M.~Andelkovic, F.~Peeters,
  \href{https://doi.org/10.5281/zenodo.4010216}{{pybinding v0.9.5: a Python
  package for tight- binding calculations}}, {This work was supported by the
  Flemish Science Foundation (FWO-Vl) and the Methusalem Funding of the Flemish
  Government.} (Aug. 2020).
\newblock \href {http://dx.doi.org/10.5281/zenodo.4010216}
  {\path{doi:10.5281/zenodo.4010216}}.
\newline\urlprefix\url{https://doi.org/10.5281/zenodo.4010216}

\bibitem{TransSIESTA}
N.~Papior, N.~Lorente, T.~Frederiksen, A.~García, M.~Brandbyge, Improvements
  on non-equilibrium and transport green function techniques: The
  next-generation transiesta, Computer Physics Communications 212 (2017) 8 --
  24.

\bibitem{Smeagol}
A.~R. Rocha, V.~M. Garc\'{\i}a-Su\'arez, S.~Bailey, C.~Lambert, J.~Ferrer,
  S.~Sanvito, Spin and molecular electronics in atomically generated orbital
  landscapes, Phys. Rev. B 73 (2006) 085414.

\bibitem{Gollum}
J.~Ferrer, C.~J. Lambert, V.~M. Garc{\'{\i}}a-Su{\'{a}}rez, D.~Z. Manrique,
  D.~Visontai, L.~Oroszlany, R.~Rodr{\'{\i}}guez-Ferrad{\'{a}}s, I.~Grace,
  S.~W.~D. Bailey, K.~Gillemot, H.~Sadeghi, L.~A. Algharagholy, {GOLLUM}: a
  next-generation simulation tool for electron, thermal and spin transport, New
  Journal of Physics 16~(9) (2014) 093029.

\bibitem{ase}
A.~H. Larsen, J.~J. Mortensen, J.~Blomqvist, I.~E. Castelli, R.~Christensen,
  M.~Du{\l}ak, J.~Friis, M.~N. Groves, B.~Hammer, C.~Hargus, E.~D. Hermes,
  P.~C. Jennings, P.~B. Jensen, J.~Kermode, J.~R. Kitchin, E.~L. Kolsbjerg,
  J.~Kubal, K.~Kaasbjerg, S.~Lysgaard, J.~B. Maronsson, T.~Maxson, T.~Olsen,
  L.~Pastewka, A.~Peterson, C.~Rostgaard, J.~Schi{\o}tz, O.~Schütt,
  M.~Strange, K.~S. Thygesen, T.~Vegge, L.~Vilhelmsen, M.~Walter, Z.~Zeng,
  K.~W. Jacobsen, The atomic simulation environment{\textemdash}a python
  library for working with atoms, Journal of Physics: Condensed Matter 29~(27)
  (2017) 273002.

\bibitem{gpaw}
J.~Enkovaara, C.~Rostgaard, J.~J. Mortensen, J.~Chen, M.~Du{\l}ak, L.~Ferrighi,
  J.~Gavnholt, C.~Glinsvad, V.~Haikola, H.~A. Hansen, H.~H. Kristoffersen,
  M.~Kuisma, A.~H. Larsen, L.~Lehtovaara, M.~Ljungberg, O.~Lopez-Acevedo, P.~G.
  Moses, J.~Ojanen, T.~Olsen, V.~Petzold, N.~A. Romero, J.~Stausholm-M{\o}ller,
  M.~Strange, G.~A. Tritsaris, M.~Vanin, M.~Walter, B.~Hammer, H.~Häkkinen,
  G.~K.~H. Madsen, R.~M. Nieminen, J.~K. N{\o}rskov, M.~Puska, T.~T. Rantala,
  J.~Schi{\o}tz, K.~S. Thygesen, K.~W. Jacobsen, Electronic structure
  calculations with {GPAW}: a real-space implementation of the projector
  augmented-wave method, Journal of Physics: Condensed Matter 22~(25) (2010)
  253202.

\bibitem{tensorflow}
M.~Abadi, A.~Agarwal, P.~Barham, E.~Brevdo, Z.~Chen, C.~Citro, G.~S. Corrado,
  A.~Davis, J.~Dean, M.~Devin, S.~Ghemawat, I.~Goodfellow, A.~Harp, G.~Irving,
  M.~Isard, Y.~Jia, R.~Jozefowicz, L.~Kaiser, M.~Kudlur, J.~Levenberg,
  D.~Man\'{e}, R.~Monga, S.~Moore, D.~Murray, C.~Olah, M.~Schuster, J.~Shlens,
  B.~Steiner, I.~Sutskever, K.~Talwar, P.~Tucker, V.~Vanhoucke, V.~Vasudevan,
  F.~Vi\'{e}gas, O.~Vinyals, P.~Warden, M.~Wattenberg, M.~Wicke, Y.~Yu,
  X.~Zheng, \href{http://tensorflow.org/}{{TensorFlow}: Large-scale machine
  learning on heterogeneous systems}, software available from tensorflow.org
  (2015).
\newline\urlprefix\url{http://tensorflow.org/}

\bibitem{kd-tree}
J.~L. Bentley, Multidimensional binary search trees used for associative
  searching, Commun. ACM 18~(9) (1975) 509--517.

\bibitem{Maneewongvatana}
S.~Maneewongvatana, D.~M. Mount, Analysis of approximate nearest neighbor
  searching with clustered point sets, in: Data Structures, Near Neighbor
  Searches, and Methodology, 1999, pp. 105--124.

\bibitem{Slater}
J.~C. Slater, G.~F. Koster, Simplified lcao method for the periodic potential
  problem, Phys. Rev. 94 (1954) 1498--1524.

\bibitem{Podolskiy}
A.~V. Podolskiy, P.~Vogl, Compact expression for the angular dependence of
  tight-binding hamiltonian matrix elements, Phys. Rev. B 69 (2004) 233101.

\bibitem{cap_model}
M.~V. Klymenko, F.~Remacle, Quantum dot ternary-valued full-adder: Logic
  synthesis by a multiobjective design optimization based on a genetic
  algorithm, Journal of Applied Physics 116~(16) (2014) 164316.

\bibitem{Baker}
A.~H. Baker, E.~R. Jessup, T.~Manteuffel, A technique for accelerating the
  convergence of restarted gmres, SIAM Journal on Matrix Analysis and
  Applications 26~(4) (2005) 962--984.

\bibitem{WIMMER20098548}
M.~Wimmer, K.~Richter, Optimal block-tridiagonalization of matrices for
  coherent charge transport, Journal of Computational Physics 228~(23) (2009)
  8548 -- 8565.

\bibitem{cormen2009introduction}
T.~Cormen, C.~Leiserson, R.~Rivest, C.~Stein, Introduction to Algorithms,
  Computer science, MIT Press, 2009.

\bibitem{Nanonet}
Nanonet: extendable python framework for the electronic structure computations
  based on the tight-binding method, \url{https://github.com/freude/NanoNet}
  (2018).

\bibitem{Jancu}
J.-M. Jancu, R.~Scholz, F.~Beltram, F.~Bassani, Empirical ${\mathrm{spds}}^{*}$
  tight-binding calculation for cubic semiconductors: General method and
  material parameters, Phys. Rev. B 57 (1998) 6493--6507.

\bibitem{Liu}
Y.~Liu, R.~E. Allen, Electronic structure of the semimetals bi and sb, Phys.
  Rev. B 52 (1995) 1566--1577.

\bibitem{laug}
E.~Anderson, Z.~Bai, C.~Bischof, S.~Blackford, J.~Demmel, J.~Dongarra,
  J.~Du~Croz, A.~Greenbaum, S.~Hammarling, A.~McKenney, D.~Sorensen, {LAPACK}
  Users' Guide, 3rd Edition, Society for Industrial and Applied Mathematics,
  Philadelphia, PA, 1999.

\bibitem{Jensen}
A.~Jensen, M.~Strange, S.~Smidstrup, K.~Stokbro, G.~C. Solomon, M.~G. Reuter,
  Complex band structure and electronic transmission eigenchannels, The Journal
  of Chemical Physics 147~(22) (2017) 224104.

\bibitem{Osca2019}
J.~Osca, L.~Serra, Complex band-structure analysis and topological physics of
  majorana nanowires, The European Physical Journal B 92~(5) (2019) 101.

\bibitem{PhysRevB.25.3975}
Y.-C. Chang, J.~N. Schulman, Complex band structures of crystalline solids: An
  eigenvalue method, Phys. Rev. B 25 (1982) 3975--3986.

\bibitem{Wimmer}
M.~Wimmer, Quantum Transport in Nanostructures: From Computational Concepts to
  Spintronics in Graphene and Magnetic Tunnel Junctions, Dissertationsreihe der
  Fakult{\"a}t f{\"u}r Physik der Universit{\"a}t Regensburg, Univ.-Verlag
  Regensburg, 2009.

\bibitem{Klymenko}
M.~V. Klymenko, J.~A. Vaitkus, J.~H. Cole, Probing charge carrier movement in
  organic semiconductor thin films via nanowire conductance spectroscopy, ACS
  Applied Electronic Materials 1~(8) (2019) 1667--1677.

\bibitem{Landauer}
Y.~Meir, N.~S. Wingreen, Landauer formula for the current through an
  interacting electron region, Phys. Rev. Lett. 68 (1992) 2512--2515.

\bibitem{FAN201428}
Z.~Fan, A.~Uppstu, T.~Siro, A.~Harju, Efficient linear-scaling quantum
  transport calculations on graphics processing units and applications on
  electron transport in graphene, Computer Physics Communications 185~(1)
  (2014) 28 -- 39.

\bibitem{KITE}
S.~M. Joao, M.~Andelkovic, L.~Covaci, T.~G. Rappoport, J.~M. V.~P. Lopes,
  A.~Ferreira, Kite: high-performance accurate modelling of electronic
  structure and response functions of large molecules, disordered crystals and
  heterostructures, Royal Society Open Science 7~(2) (2020) 191809.
\newblock \href {http://dx.doi.org/10.1098/rsos.191809}
  {\path{doi:10.1098/rsos.191809}}.

\bibitem{KPM}
A.~Wei\ss{}e, G.~Wellein, A.~Alvermann, H.~Fehske, The kernel polynomial
  method, Rev. Mod. Phys. 78 (2006) 275--306.

\bibitem{TBTK}
K.~Bjornson, Tbtk: A quantum mechanics software development kit, SoftwareX 9
  (2019) 205 -- 210.

\bibitem{PhysRevB.85.035317}
G.~Mil'nikov, N.~Mori, Y.~Kamakura, Equivalent transport models in atomistic
  quantum wires, Phys. Rev. B 85 (2012) 035317.

\bibitem{PhysRev.137.A871}
L.~M. Falicov, S.~Golin, Electronic band structure of arsenic. i.
  pseudopotential approach, Phys. Rev. 137 (1965) A871--A882.

\end{thebibliography}

\end{document}